\documentclass[twocolumn,showpacs,preprintnumbers,aps,pra,amsmath,amssymb,floatfix]{revtex4}
\usepackage{graphicx}
\usepackage{dcolumn}
\usepackage{bm}
\usepackage{amssymb}
\usepackage{setspace}

\begin{document}

\title{Coherent control of ultracold $^{85}$Rb trap-loss collisions with nonlinearly frequency-chirped light}

\author{J. A. Pechkis$^{1,2}$, J. L. Carini$^1$, C. E. Rogers III$^1$,  and P. L. Gould$^1$}

\affiliation{$^1$Department of Physics, University of Connecticut,
Storrs, CT 06269, USA\\
$^2$Currently with the Naval Research Laboratory, 4555 Overlook Avenue S.W., Washington, DC 20375, USA}

\author{S. Kallush$^3$ and R. Kosloff$^4$}
\affiliation{$^3$Department of Physics and Optical Engineering, ORT Braude, P.O. Box 78, Karmiel , Israel\\$^4$Department of Physical Chemistry and the Fritz Haber Research Center for Molecular Dynamics, The Hebrew University, 91094, Jerusalem, Israel}

\date{\today}

\begin{abstract}

We present results on coherent control of ultracold trap-loss collisions using 40 ns pulses of nonlinearly frequency-chirped light. The chirps, either positive or negative, sweep $\sim$ 1 GHz in 100 ns and are centered at various detunings below the $D_2$ line of $^{85}$Rb. At each center detuning, we compare the collisional rate constant $\beta$ for chirps that are linear in time, concave-down and concave-up. For positive chirps, we find that $\beta$ generally depends very little on the shape of the chirp. For negative chirps, however, we find that $\beta$ can be enhanced by up to 50(20)$\%$ for the case of the concave-down shape. This occurs at detunings where the evolution of the wavepacket is expected to be coherent. An enhancement at these detunings is also seen in quantum mechanical simulations of the collisional process.
\end{abstract}

\pacs{32.80.Qk, 32.80.Pj, 34.50.Rk}

\maketitle

\section{Introduction}

In recent years there has been increasing interest in applying the techniques of coherent control to ultracold systems. Coherent control \cite{Rice00,Brumer03} generally involves the use of shaped laser pulses on ultrafast timescales to control internal degrees of freedom, while ultracold physics \cite{Metcalf99} deals with external degrees of freedom at sub-millikelvin temperatures. One topic where these fields can productively overlap is the formation of ultracold molecules by coherently-controlled photoassociation of ultracold atoms \cite{LucBook}. Since ultracold molecules \cite{Dulieu,Carr,Faraday,Krems} have many potential applications in quantum information, ultracold chemistry, dipolar systems, and precision spectroscopy, it is important to develop techniques for their efficient and controlled production. In the present work, we investigate a process closely related to photoassociation: long-range excited-state collisions between ultracold atoms. In particular, we demonstrate an improved coherent control of these inelastic collisions using shaped frequency chirps on the nanosecond timescale. This is relevant to photoassociative molecule formation because the bottleneck for this process is often the paucity of atom pairs at the desired internuclear separation $R$.

\begin{figure}
\centerline{\includegraphics[width=0.98\linewidth]{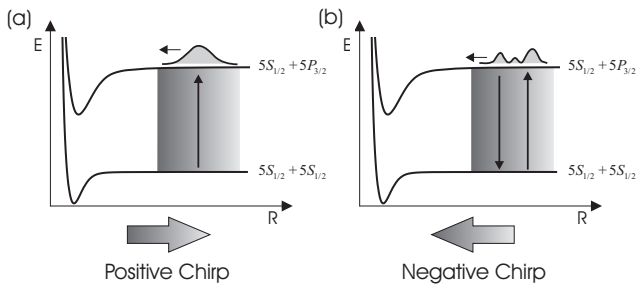}}
\caption{Ultracold collisions induced by frequency-chirped light. The ground- and excited-state potentials are shown as well as the excited-state wavepacket and the region of $R$ swept over by the chirp. (a) Positive chirp. The excitation radius (upward arrow) moves outward with time, while the excited-state wavepacket moves inward. (b) Negative chirp. The excitation radius moves inward in time, following the excited wavepacket trajectory, leading to multiple interactions, which can return a portion of the wavepacket to the ground state (downward arrow).}
\label{fig:fig_1}
\end{figure}

Although there has been much theoretical work aimed at coherently-controlled photoassociative formation of ultracold molecules \cite{Vala00,Luc04,LucII04,Koch06,Poschinger06,KochII06,Brown06,KochIII06,Petit,KallushI,KallushII,KallushIII,Koch09}, initial experiments demonstrated only the coherent control of the photodestruction of ultracold molecules using shaped \cite{Salzmann06} and chirped \cite{BrownII06} ultrafast pulses. Subsequent experiments \cite{Salzmann08,Mullins09,Merli09,McCabe09,Martay09} observed coherent transients in excited-state molecules photoassociated with femtosecond pulses and there was some evidence for the production of ground-state molecules \cite{Mullins09}. In contrast, our approach is to control the dynamics of ultracold atoms at large $R$ using nanosecond-timescale frequency-chirped pulses. Coherent control of these long-range interactions, and the subsequent evolution of the atom pair to short range, may allow improved photoassociative formation of ultracold molecules. In our previous work, we demonstrated that such linearly chirped pulses can efficiently excite Rb atom pairs over a wide range of R, leading to ultracold trap-loss collisions \cite{Wright05}. Multiple-pulse effects were also investigated. We found that the rate of collisions induced by a given  pulse can be enhanced or suppressed by a preceding pulse, depending on the delay \cite{Wright06}. In work most relevant to that described here, we demonstrated coherent control of these ultracold trap-loss collisions \cite{Wright07}. Specifically, we observed significantly different behaviors for positive and negative linear chirps. The negative chirp is of particular interest because its radius for resonant excitation to the attractive molecular potential moves inward with time, in the same direction as the excited-state wavepacket. This allows multiple interactions between the chirped light and the excited atom pair (see Fig. \ref{fig:fig_1}(b)). A possible outcome is coherent collision blocking, where the light coherently de-excites the wavepacket at long range, thereby suppressing the rate of trap-loss collisions. The experiments described here extend our earlier work by exerting a further degree of control over the collisional process. Our previous control parameter was simply the sign of the linear chirp. Here we actually change the shape of the chirp by making it nonlinear. In the regime where coherent collision blocking takes place, we find a significant dependence on the details of this nonlinearity. Quantum mechanical simulations of the collisional process show a similar dependence.

The physics of the trap-loss collisions induced by frequency-chirped light is shown schematically in Fig. 1. For a given detuning $\Delta$ ($\Delta<0$) below the atomic resonance, the photon energy matches the energy difference between the ground-state molecular potential, assumed flat, and the long-range attractive excited-state potential, $-C_3/R^3,$ at the Condon radius $R_C = \left[-C_3/(\hbar \Delta)\right]^{1/3}$. As the laser frequency is chirped, this excitation radius changes with time, sweeping inward (outward) for a negative (positive) chirp. After excitation, the excited-state wavepacket quickly accelerates inward where it can gain sufficient kinetic energy (e.g., by fine-structure predissociation from a short-range curve crossing \cite{Julienne91}) to escape the trap. The rate constant $\beta$ for these trap-loss collisions depends not only on the center frequency of the chirped pulse, $\Delta_\mathrm{p}$, but also on the direction of the chirp \cite{Wright07}, and as we show below, on the nonlinearity of the frequency chirp. The dependence on chirp direction arises from the fundamental asymmetry in the system: while the excitation radius can move either inward or outward with time, the excited-state wavepacket always moves inward. Therefore, the negative chirp can provide multiple interactions, while the positive chirp only interacts once. The dependence on nonlinearity arises from the details of the excited-state wavepacket evolution.

\section{Experiment}

We measure the collisional trap-loss rate constant $\beta$ via the density-dependent decay of a sample of ultracold ($\sim$ 50 $\mu$K) $^{85}$Rb atoms in a magneto-optical trap (MOT) \cite{Wright05}. Excited-state trap-loss collisions are induced by  40 ns full-width-at-half-maximum (FWHM) frequency-chirped Gaussian pulses. The MOT light is turned off for 150 $\mu$s, during which time a number (typically 60) of frequency-chirped pulses with peak intensity $I=67$ W/cm$^2$ illuminates the atoms at a repetition rate of 1 MHz. This 1 $\mu$s delay between pulses ensures that each chirped pulse acts independently \cite{Wright06}. The overall sequence is repeated every 722 $\mu$s. The MOT repump light remains on throughout the cycle to correct for any optical pumping induced by the frequency-chirped light.

\begin{figure}
\centerline{\includegraphics[width=0.98\linewidth]{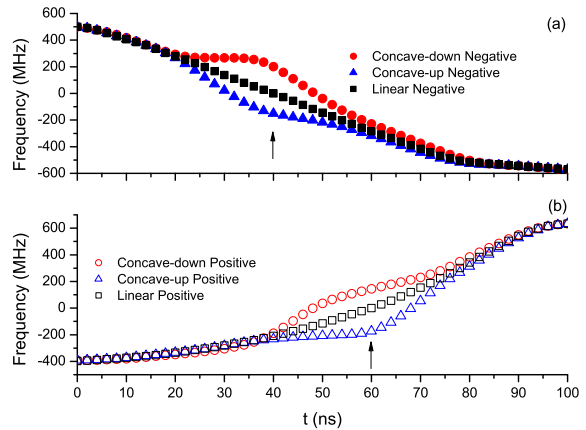}}
\caption{(color online) Smoothed interpolations of the nonlinear frequency chirps (concave-down, concave-up, and linear), extracted from optical heterodyne measurements for both (a) negative and (b) positive chirps. The frequencies are relative to the centers of the linear chirps. The chirps are shaped over an interval corresponding to the 40 ns FWHM of the Gaussian pulse used to select out the chirp. During this time, the slope is $<$ 0 for negative chirps or $>$ 0 for positive chirps, with a frequency difference of $\sim$ 400 MHz between concave-down and -up chirps at the peak intensity of the pulse (indicated by the arrow).}
\label{fig:fig_2}
\end{figure}

The frequency-chirped light, which sweeps approximately 1 GHz in 100 ns, is produced by rapidly modulating the current of external-cavity diode laser (ECDL) with a 240 MHz arbitrary waveform generator (Tektronix AFG 3252) \cite{Wright04}. The output of the ECDL injection-locks a free-running diode laser, and an acousto-optical modulator selects out the desired portion of the frequency-chirp. Smoothed interpolations of the nonlinear frequency chirps, extracted from optical heterodyne measurements, are shown in Fig. \ref{fig:fig_2}. We restrict the shaping of the nonlinear chirps (labeled as concave-down, concave-up, and linear, based on the shapes of the respective frequency plots) to the duration corresponding to the 40 ns FWHM of the Gaussian pulse, which is centered at 40 ns (60 ns) during the negative (positive) chirp. This ensures that the various nonlinear chirps sweep over the same effective chirp range. Furthermore, the frequency chirps are tailored such that their slopes always remain either negative or positive. The maximum frequency difference between concave-down and concave-up chirps is $\sim$ 400 MHz. This occurs at the peak intensity of the pulse. 

\section{Results and Analysis}
\label{sec:results}

\subsection{Experimental results}
\label{sec:exp}

\begin{figure}[bt]
  \centering
  \includegraphics[width=0.9\linewidth]{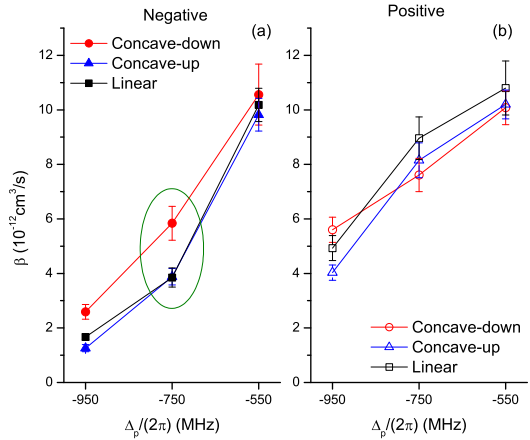}
  \caption{(color online) (a) $\beta(\Delta$$_\mathrm{p}$) for concave-down, concave-up, and linear negative chirps. In the region of $\Delta_\mathrm{p}$ where coherent collision blocking is important, we find a dependence of $\beta$ on the details of the nonlinearity of the negative chirp. In particular, the negative concave-down chirp yields a higher $\beta$ than those of the linear and concave-up negative chirps for $\Delta_\mathrm{p}$/(2$\pi$) = -750 MHz (circled). (b) $\beta(\Delta_\mathrm{p})$ for concave-down, concave-up, and linear positive chirps. We find no significant dependence of $\beta$ on the nonlinearity, except possibly at -950 MHz. This is consistent with the fact that the excited-state wavepacket moves inward on the attractive potential while the Condon radius associated with the frequency chirp sweeps outwards.}
  \label{fig:fig_3}
\end{figure}

We measure $\beta$ as a function of the center detuning $\Delta_\mathrm{p}$ (with respect to the $F=3\rightarrow F'=4$ transition) of our frequency-chirped pulse in the region of detunings where coherent collision blocking takes place. Our results comparing the various nonlinear negative chirps are shown in Fig. \ref{fig:fig_3}(a). As can be seen from the data, there is a dependence of $\beta$ on the nonlinearity of the negative chirp. We find that the concave-down negative chirp yields a value of $\beta$ that is 50(20)$\%$ larger that those of the linear and concave-up chirps at $\Delta_\mathrm{p}/(2\pi)$ = -750 MHz. A similar behavior is seen at $\Delta_\mathrm{p}/(2\pi)$ = -950 MHz. This difference is not due simply to the fact that the frequency at which the pulse intensity peaks is different for the various chirp shapes. The linear and concave-up negative chirps yield similar values of $\beta$, supporting the idea that the details of the frequency chirp shape are important. Since coherent collision blocking is important in this region, this is a demonstration of the coherent control of the collisional process through shaping of the frequency chirp. We should not be surprised that our simple shaping does not yield as large a difference (a factor of 2.3) in $\beta$ as is seen when comparing the positive and negative linear chirps at $\Delta_\mathrm{p}/(2\pi)$ = -750 MHz, since reversing the sign of the chirp is a more extreme change than the concavity added to the negative chirps.

At $\Delta_\mathrm{p}/(2\pi)$ = -550 MHz, we believe that incoherent flux enhancement starts to become more important, resulting in an increase of $\beta$. This flux enhancement is due to spontaneous emission, followed by re-excitation at shorter range, during the chirp. It was previously observed, in both the data and in classical Monte-Carlo simulations, for similar detunings \cite{Wright07}.

We compare the above results to those of the shaped positive chirps, which are shown in Fig. \ref{fig:fig_3}(b). As can be seen from the data, we find, as expected, no significant difference in $\beta$ due to shaping the positive chirps, except possibly at $\Delta_\mathrm{p}/(2\pi)$ = -950 MHz. Again, this is due to the fact that the excitation (Condon) radius associated with the positive chirp sweeps away from the evolving excited-state wavepacket. This is important because it verifies that the effective chirp range seen by the atom-pair is the same for positive concave-down, concave-up, and linear chirps. Therefore, we can confidently say that the difference in $\beta$ observed for negative chirps in the region where coherent collision blocking takes place is solely due to the details of the nonlinearity of the negative chirp. The values of beta increase for less negative detunings because there are more atom pairs available for excitation at the corresponding larger values of $R$.

\begin{figure}[bt]
  \centering
  \includegraphics[width=0.9\linewidth]{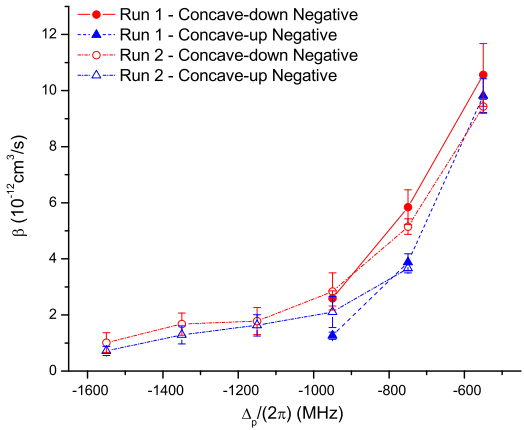}
  \caption{(color online) $\beta(\Delta$$_\mathrm{p}$) for concave-down and concave-up negative chirps for large $\Delta_\mathrm{p}$. An additional set of data was taken for negative concave-down and concave-up chirps for -1550 MHz $\leq\Delta_\mathrm{p}\leq$ -550 MHz (Run 2) to determine the behavior of $\beta$ at large red detunings. When this data is multiplied by a factor of 1.55, it agrees well with the nonlinear negative chirp data previously presented (Run 1). For $\Delta_\mathrm{p}<$ -950 MHz, there is no difference in $\beta$($\Delta_\mathrm{p}$) between negative concave-down and -up chirps. This is consistent with the fact that the dynamics of the excited-state wavepacket evolve too quickly for the shaped negative chirps to follow. Multiple interactions are thereby avoided.}
  \label{fig:fig_4}
\end{figure}

Finally, we examine the effect of the nonlinearity of the negative chirp on $\beta$ for larger negative detunings. For this data set, $\beta$ is measured for concave-down and concave-up negative chirps for -550 MHz $\leq \Delta_\mathrm{p}/(2\pi) \leq$ -1550 MHz as shown in Fig. \ref{fig:fig_4} where the data are scaled by 1.55 to match with the data presented earlier. This scaling is consistent with the fact that absolute values of $\beta$ are uncertain within a factor of $\sim$ 2 due to uncertainties in the atomic density calibration which can vary from run to run. With this scaling, the two data runs agree rather well for $\Delta_\mathrm{p}/(2\pi) \geq$ -950 MHz. Again, we see a difference in $\beta$ for the two chirp shapes at $\Delta_\mathrm{p}/(2\pi)$ = -750 MHz: the negative concave-down chirp yields a larger $\beta$ than that of the negative concave-up chirp.

For $\Delta_\mathrm{p}/(2\pi) <$ -950 MHz, we find that the concave-down and concave-up negative chirps yield the same dependence of $\beta$ on $\Delta_\mathrm{p}$. Specifically, values of $\beta$ for the two chirp shapes converge for $\Delta_\mathrm{p}/(2\pi) <$ -950 MHz, and they both decrease with larger negative values of $\Delta_\mathrm{p}$. This same phenomenon was observed in the data and simulations (both classical and quantum mechanical) for positive and negative linear chirps \cite{Wright07}. This is due to the fact that the wavepacket accelerates so quickly on the steep potential that the negative frequency chirp cannot $''$catch up$''$ with it after the initial excitation. This mismatch of time scales between the chirp and the internuclear dynamics yields a rate of trap-loss collisions that is independent of the sign of the chirp. For the same reason, the shaped chirps used here yield the same collisional loss rates for large red detunings. This is also further evidence that the effective chirp range seen by the atom pair is the same for the concave-down and concave-up chirps. Also, in agreement with our previous observations for linear positive and negative chirps \cite{Wright07}, $\beta$ decreases as the center detuning becomes more negative. This is because the excitation is occurring at shorter range, where less atom pairs are available. 

\subsection{Quantum mechanical simulations}
\label{sec:sim}

\begin{figure}
\centering
\includegraphics[width=0.95\linewidth]{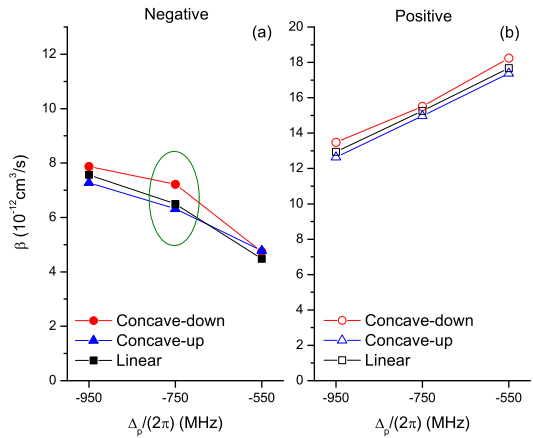}
\caption{(color online) Results of quantum mechanical simulations of $\beta(\Delta$$_\mathrm{p}$) for nonlinear negative (a) and positive (b) chirps. The largest variation with shape occurs for the negative chirp at -750 MHz (circled), as in the experiment (Fig. \ref{fig:fig_3}(a)).}
\label{fig:fig_5}
\end{figure}

The theory behind the quantum mechanical simulations is described in detail in Ref. \cite{Wright07} and references therein, and the results of the simulations for the present experiment (shown in Fig. \ref{fig:fig_5}) and their interpretation will be described in more detail in Ref. \cite{Carini10}. Briefly, the time-dependent Schr\"{o}dinger equation is solved numerically for the ground and excited (0$_u^+$) internuclear radial wavefunctions of an atom pair in the presence of the chirped laser pulse. The initial state is assumed to be the zero-energy $s$-wave scattering state for either the singlet or triplet ground state. Any excited-state flux crossing a short-range ($R=100$ $a_0$) absorbing boundary is assumed to give rise to an inelastic trap-loss collision. Spontaneous emission is included as an artificial sink channel for the excited state. The model therefore does not allow for re-excitation following spontaneous emission (e.g., flux enhancement) or incoherent high-field effects. The overall trap-loss rate constant is taken as a weighted average over the singlet and triplet ground states.

Focusing on the results of the quantum mechanical simulations for the nonlinear negative chirps first (Fig. \ref{fig:fig_5}(a)), we see that at $\Delta_\mathrm{p}/(2\pi)$ = -550 MHz, the various chirps yield the same $\beta$, which is the case in the experiment. However, flux enhancement is not included in the quantum mechanical simulations \cite{Carini10}, and therefore, $\beta$ at -550 MHz is lower than the values for $\Delta_\mathrm{p}/(2\pi)$ $\leq$ -750 MHz. At $\Delta_\mathrm{p}/(2\pi)$ = -750 MHz, $\beta_{\mathrm{down}}^{\mathrm{neg}} > \beta_{\mathrm{up}}^{\mathrm{neg}} \simeq \beta_{\mathrm{lin}}^{\mathrm{neg}}$ for concave-down, concave-up, and linear chirps, respectively, as is seen in the experiment. From the dynamics provided by the quantum mechanical simulations \cite{Carini10}, we see that a larger portion of the wavepacket created by the concave-down chirp is able to more quickly accelerate on the excited-state potential than the wavepackets of the concave-up and linear chirps near the beginning of the shaped portions of the chirps. It appears that this decreased ability to match the excited-state wavepacket dynamics may lead to a larger $\beta$ for the concave-down chirp. For $\Delta_\mathrm{p}/(2\pi)$ = -950 MHz, the values of $\beta$ for the various chirp shapes begin to converge, although the trend seen at -750 MHz is maintained.

The results of quantum mechanical simulations for the nonlinear positive chirps, shown in Fig \ref{fig:fig_5}(b), yield values of $\beta$ for the various chirp shapes that are similar at a given $\Delta_\mathrm{p}$. Therefore, there is little dependence of $\beta$ on the details of the nonlinearity of the positive chirp, as expected, and as seen in the experiment. These results for the nonlinear positive chirps are consistent with the fact that the excitation radius of the positive chirp sweeps outward, away from the excited-state wavepacket trajectory. As noted previously \cite{Wright07}, we see that the positive chirp generally gives larger values of $\beta$ than the negative chirp, both in the data (Fig. \ref{fig:fig_3}) and in the simulations (Fig. \ref{fig:fig_5}). This is attributed to efficient adiabatic excitation, with no further interactions, for the case of the positive chirp.

\section{Conclusion}

We have shown experimental results that demonstrate coherent control of atomic trap-loss collisions with nonlinearly frequency-chirped light on the nanosecond timescale. In particular, we find a dependence of the collisional loss rate constant $\beta$ on the details of the nonlinearity of the negative chirp. For center detunings $\Delta_\mathrm{p}$ where coherent collision blocking takes place, the concave-down negative chirp yields higher values of $\beta$ than those of the concave-up and linear negative chirps. For more negative $\Delta_\mathrm{p}$, which corresponds to excitation at shorter range, $\beta$ for the negative chirp becomes independent of the nonlinearity because the excited-state wavefunction accelerates more rapidly on the steep attractive potential. In general, for positive chirps we find no significant dependence of $\beta$ on the nonlinearity. Our experimental results are supported by quantum mechanical simulations of the collisional process. Together, the experiment and the simulations highlight the importance of matching the chirp and the excited-state wavefunction evolution. Our work demonstrates that the shaping of frequency-chirped pulses on the nanosecond timescale is a viable technique for coherently controlling interactions between ultracold atoms. Since a negative chirp can excite a wavepacket at long range and de-excite it at short range, these coherent interactions may allow enhancement of photoassociative molecule formation. Our recent observation of ultracold molecules formed using frequency-chirped pulses \cite{JPechkis_Thesis} is encouraging in this regard. The use of faster and better-controlled chirps \cite{Rogers07} will facilitate these efforts.

\begin{acknowledgments}
The work at the University of Connecticut is supported by the Chemical Sciences, Geosciences and Biosciences Division, Office of Basic Energy Sciences, Office of Science, U.S. Department of Energy.
\end{acknowledgments}

\end{document}